# Statistical-mechanical lattice models for protein-DNA binding in chromatin

**Vladimir B Teif[1,2] and Karsten Rippe[1]**

[1]Research Group Genome Organization & Function, Deutsches Krebsforschungszentrum and BioQuant, Im Neuenheimer Feld 280, 69120 Heidelberg, Germany;
[2]Institute of Bioorganic Chemistry, Belarus National Academy of Sciences, Kuprevich 5/2, 220141, Minsk, Belarus.

E-mail: Vladimir.Teif@bioquant.uni-heidelberg.de

**Abstract.** Statistical-mechanical lattice models for protein-DNA binding are well established as a method to describe complex ligand binding equilibriums measured *in vitro* with purified DNA and protein components. Recently, a new field of applications has opened up for this approach since it has become possible to experimentally quantify genome-wide protein occupancies in relation to the DNA sequence. In particular, the organization of the eukaryotic genome by histone proteins into a nucleoprotein complex termed chromatin has been recognized as a key parameter that controls the access of transcription factors to the DNA sequence. New approaches have to be developed to derive statistical mechanical lattice descriptions of chromatin-associated protein-DNA interactions. Here, we present the theoretical framework for lattice models of histone-DNA interactions in chromatin and investigate the (competitive) DNA binding of other chromosomal proteins and transcription factors. The results have a number of applications for quantitative models for the regulation of gene expression.

## 1. Introduction

Recent breakthroughs in high-throughput DNA sequencing technologies as an approach to identify protein-associated DNA fragments have created the unprecedented situation that genome-wide experimental data are being accumulated faster than the development of the corresponding biophysical models for gene regulation (Jiang and Pugh, 2009; Segal and Widom, 2009a; Radman-Livaja and Rando, 2009; Cairns, 2009). In eukaryotes, transcription factors (TF) do not interact with free DNA, but rather compete with complex multi-component assemblies of histones and other chromosomal proteins for DNA-binding sites. The elementary unit of chromatin is the nucleosome. It consists of a histone octamer complex containing two copies of each histone H2A, H2B, H3 and H4, around which 146-147 base pairs of DNA are wrapped in 1.67 turns. Nucleosomes are connected via the intervening linker DNA and the chain of nucleosomes folds into a chromatin fiber. Both the position of the nucleosomes as well as the fiber organization determine DNA accessibility by TFs and RNA polymerases (Boeger *et al.*, 2008; Kim and O'Shea, 2008; Lam *et al.*, 2008; Petesch and Lis, 2008; Gilmour, 2009). Depending on the context, nucleosomes can inhibit (Whitehouse *et al.*, 2007; Henikoff, 2008) or facilitate (Zhao *et al.*, 2001) transcription factor binding. Nucleosome positions are controlled by three major contributions: First, the intrinsic binding affinity of the histone octamer depends on the DNA sequence (Thåström *et al.*, 1999; Segal *et al.*, 2006; Whitehouse *et al.*, 2007; Whitehouse and Tsukiyama, 2006; Ioshikhes *et al.*, 2006; Peckham *et al.*, 2007). Second, the nucleosome can be displaced or recruited by the competitive or cooperative binding of other protein factors (Wasson and Hartemink, 2009; Workman and Kingston, 1992; Morozov *et al.*, 2009). Third, the nucleosome may be actively translocated by ATP-dependent remodeling complexes (Teif and Rippe, 2009; Zhang *et al.*, 2009a; Schnitzler, 2008; Radman-Livaja and Rando, 2009; Whitehouse and Tsukiyama, 2006; Hartley and Madhani, 2009; Rippe *et al.*, 2007).

Most current theoretical descriptions assume that gene-regulatory events are governed by mass-action laws and can be approximated by a thermodynamic equilibrium description (Bintu *et al.*, 2005a; Bintu *et al.*, 2005b). Accordingly, DNA occupancies of transcription factors and RNA polymerase are related to gene expression probabilities as functions of protein concentrations (Teif, 2010). Several approaches have been reported to predict gene expression from the DNA sequence using these concepts with promising results, but currently the contribution of the chromatin organization is not accounted for (Beer and Tavazoie, 2004; Segal *et al.*, 2008). Since DNA is characterized by the sequence of its constituting base pairs, it can be represented by a one-dimensional lattice. The complexity of the associated protein interaction network in chromatin is reflected in different types of models. These are usually constructed to investigate the relation between protein composition and DNA sequence, localization of proteins and the biological activity of chromatin in terms of gene expression. The most basic description would comprise only the linker DNA and the histone octamer occupying binding sites of 147 bps of the DNA chain in the nucleosome. The nucleosome position can be changed by energy-dependent molecular motors referred to as chromatin remodelers (see (Teif and Rippe, 2009) and references therein). The models might consider the existence of subnucleosomal particles in which not all eight histones are present, the linker histones H1 that can organize an additional 20 bp of the linker DNA at the nucleosome, and/or the exchange of canonical core histones with variants (reviewed in (Rippe *et al.*, 2008)). In addition to histones, other chromosomal proteins such as HMG and heterochromatin protein 1 (HP1) may assemble along the DNA lattice and act as important architectural components for the folding of the chain. Finally, gene specific transcription factors may display competing or cooperative binding with the above chromatin protein components at regulatory DNA regions such as promoters or enhancers. Thus, depending on the questions to be addressed, specific lattice models need to be devised. Here we define a theoretical framework that can be adapted to account for all the above scenarios, and apply it to the interdependence of histone and TF binding to DNA in the context of gene regulation.

## 2. Lattice models for DNA-protein binding

*2.1. General considerations*
Chromatin has a complex three-dimensional organization. The DNA is wrapped around the histone octamer core and the resulting nucleosome chain can fold into a variety of conformations (Kepper *et al.*, 2008). However, all protein binding events can be assigned to the linear sequence of DNA base pairs, which allows constructing a one-dimensional representation of the binding sites. The classical principles of calculation of macromolecule binding to a one-dimensional DNA lattice were formulated several decades ago (Poland, 1979; Hill, 1985). In mathematics, this field is known as "sequential adsorption" as well as "car parking" problems (Evans, 1993). In computational biology such models are formulated with the help of Markov chains (e.g. Hidden Markov Models, HMM), which go back to the works of the mathematician Andrei Markov (Markov, 1907). In biophysics they are usually known as Ising models historically arising from the work of Ernst Ising on the theory of ferromagnetism (Ising, 1925). The general concept is to divide the system into the elementary units that are associated with different states (e.g. bound/unbound). The states of the whole system are then calculated as different combinations of states of elementary units. The elementary DNA units are usually taken as nucleotides A, T, C and G that represent one DNA strand in the double helix with base-pairs $A \equiv T$, $T \equiv A$, $C \equiv G$ and $G \equiv C$. Other choices of the elementary lattice unit are also possible as discussed below.

Statistical-mechanical lattice models are usually based on the assumptions that proteins bind DNA reversibly, that binding affinities are determined by the DNA- and protein sequence, and that protein activities are proportional to their concentrations. The binding probabilities calculated for the lattice models are then equal to the fraction of molecules in a given state within the statistical ensemble consisting of many identical systems. Within a statistical ensemble, each state of the system is characterized by a weight $\exp(-\Delta G_i / k_B T)$, where $\Delta G_i$ is the energy difference of this state with respect to some reference state. In particular, when protein binding to the DNA is considered, the unbound state is usually set as a reference state with weight 1. Each bound state is then characterized by a weight $K_{ng} c_{0g}$, where $K_{ng}$ is the binding constant for the protein of type $g$ to the DNA site $n$, and $c_{0g}$ is the concentration of $g$-type protein in solution. The linear dependence on concentration follows from the dilute-solution assumption that the protein activity is proportional to the concentration. (As said above, this assumption may not hold in chromatin, which may in principle lead to nonlinear weights – the possibility which has not been tested yet). The partition function Z is equal to the sum of the weights of all possible states:

$$Z = \sum_i e^{-\Delta G_i / k_B T} \tag{1}$$

The probability $P_x$ of a given state $x$ is a ratio of the weight of a given state and the partition function $Z$:

$$P_x = \frac{e^{-\Delta G x / k_B T}}{Z} \quad (2)$$

If a weight $K_x$ corresponding to the state $x$ enters the partition function linearly, the following expression holds for the probability of this state $P_x$:

$$P_x = \frac{\partial Z}{\partial K_x} \times \frac{K_x}{Z} \quad (3)$$

In particular, the probability of binding of a protein of type $g$ to a DNA site $n$ is determined by the derivative of the partition function by the corresponding binding constant $K_{ng}$ (Teif, 2007):

$$P_{ng} = \frac{\partial Z}{\partial K_{ng}} \times \frac{K_{ng}}{Z} \quad (4)$$

One-dimensional models may be formulated and solved mathematically using different algorithms, which are more or less computationally effective. Several methods of constructing lattice models of DNA-protein binding have been developed in the past. They may be roughly divided into four classes (Table 1): direct combinatorial methods (Zasedatelev *et al.*, 1971; McGhee and von Hippel, 1974; Tsuchiya and Szabo, 1982; Nechipurenko and Gursky, 1986; Wolfe and Meehan, 1992), generating functions methods (Lifson, 1964; Schellman, 1974; Chen *et al.*, 1986; Chen, 1987; Chen, 1990), recurrent relations methods (Gurskii and Zasedatelev, 1978; Nechipurenko *et al.*, 2005; Segal *et al.*, 2006; Wasson and Hartemink, 2009; Granek and Clarke, 2005; Laurila *et al.*, 2009) and transfer matrix methods (Teif, 2007; Hill, 1957; Gurskii *et al.*, 1972; Crothers, 1968; Chen, 1987; Akhrem *et al.*, 1985; Chen, 2004; Di Cera and Kong, 1996). The difference between computational algorithms lies mostly in the way they enumerate the states of the system and assign corresponding weights, as described in more detail below. The enumeration of states also depends on the features of protein-DNA binding which are taken into account. Basic DNA-binding features include: (i) site specificity determined by the DNA sequence, (ii) overlapping binding sites, (iii) competitions between different protein types or different binding modes, (iv) interactions between proteins bound to the DNA (e.g. contact interactions in transcription complexes assembled from several subunits with "sticky ends", (v) long-range interactions over larger distances along the double helix through DNA conformational transitions), (vi) multilayer binding (when a protein bound to the DNA presents a lattice for the next-layer binding of other proteins), and (vii) protein-assisted DNA looping. The strength of the combinatorial method and the generating functions method is mostly the possibility to derive analytical solutions for simple systems such as non-specific protein binding to infinitely large DNA. However, site-specific binding requires calculations according to the real DNA sequence, which rules out analytical solutions. The recurrent relations method easily solves the problem of sequence-specificity and usually allows the fastest algorithms for realistic genomic regions. On the other hand, this method is more difficult to implement to the problems involving DNA loops and multilayer protein assembly. The transfer matrix method allows the highest flexibility to describe models of almost any degree of complexity, but the acceleration of such calculations requires additional efforts. Thus, the search for more effective computational algorithms for one-dimensional models is still an open area. Finally, once a model is defined in terms of the elementary lattice unit, its available states and statistical weights, the results of the calculation should not depend on the choice of the computational algorithm.

**Table 1. Basic features of DNA-protein binding and lattice approaches**

| Feature \ Method | Combinatorics | Generating funct. | Transfer matrices | Recurrent relations | Graphical representation |
|---|---|---|---|---|---|
| Sequence specificity | - | + | + | + | 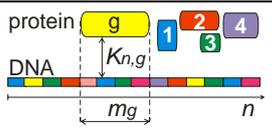 |
| Overlapping binding sites | + | + | + | + | |
| Multiprotein competition | - | - | + | + | |
| Contact interactions | + | + | + | + | 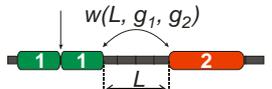 |
| Long-range interactions | + | + | + | + | |
| Multilayer binding | - | + | + | - | 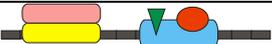 |
| Short DNA loops | - | - | + | + | 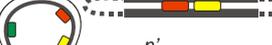 |
| Long DNA loops (e.g. promoter-enhancer) | + | - | + | - | 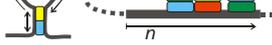 |

*2.2. Combinatorial method*

The combinatorial method uses binomial formulas to derive analytical expressions for the numbers of possible rearrangements of proteins along the DNA (McGhee and von Hippel, 1974; Zasedatelev *et al.*, 1971). For example, a protein, which covers $m$ base pairs upon binding to the DNA, may adopt ($N - m + 1$) positions along the DNA double helix of length $N$. This yields $[(N - k \times (m - 1))! / (N - m \times k)! / k!]$ possible rearrangements of $k$ proteins. For a nonspecific binding to infinitely long DNA ($N >> n$), the degree of protein binding $v = k / N$ is linked to its bulk concentration $c_0$, binding site size $m$ and binding constant $K$ according to the following relation (McGhee and von Hippel, 1974):

$$\frac{v}{c_0} = K(1-mv)\left(\frac{1-mv}{1-(m-1)v}\right)^{m-1} \tag{5}$$

Exact analytical solutions such as the McGhee-von Hippel formula are widely used for the description of *in vitro* binding experiments. More specific and more general solutions also exist, which consider cooperative interactions between proteins (Zasedatelev *et al.*, 1971), polarity of protein-protein interactions (Nechipurenko *et al.*, 1979; Wolfe and Meehan, 1992), competitions between specific and nonspecific binding on a short DNA oligomer (Tsodikov *et al.*, 2001), electrostatic interactions between the molecules upon binding to the DNA (Rouzina and Bloomfield, 1997), ligand-induced double-helix melting (Lando, 1994), adsorptive phase transitions (Lando and Teif, 2000) DNA condensation induced by binding of multivalent counterions (Maltsev *et al.*, 2006; Teif, 2005) and binding of flexible branched oligopolymers (Horsky, 2008; Nishio and Shimizu, 2005).

*2.3. Generating functions method*

The generating functions method is based on the idea to characterize the system by a mathematical expression called the generating function instead of the partition function (Lifson, 1964; Schellman, 1974; Kong, 2001). The generating function is defined by Eq. 6:

$$G(c_0, x) = \sum_{n=0}^{\infty} Z_n(c_0) x^n \qquad (6)$$

Here $G(c_0, x)$ is the generating function of the system characterized by the partition functions $Z_n(c_0)$, where $n$ is the number of DNA-bound proteins in a given configuration, and $c_0$ stands for all input concentrations (if there is only one protein type, then $c_0$ has the same meaning of a free protein concentration as in Eqs. 1-5). Variable $x$ has no direct physical meaning. The idea is that during the algebraic manipulations $x$ should vanish and infinite series converge to well-defined analytical limits. In the case of small non-interacting ligands binding to one unit of an infinite DNA lattice, the generating function is given by Eq. 7 (Kong, 2001):

$$G(c_0, x) = \frac{1}{1 - (1 + Kc_0)x} = \sum_{n=0}^{\infty} (1 + Kc_0)^n x^n \qquad (7)$$

This gives a trivial solution for the partition function with $Z = (1 + Kc_0)^n$. If a protein covers $m$ lattice units upon binding to the infinitely long DNA, the generating function method yields the McGhee-von Hippel expression (Eq. 5). This method is also applicable to more complex modes of binding. For example, it successfully allows treating the "piggy-back" binding problem, when a protein bound to the DNA can bind other proteins on its back (Chen *et al.*, 1986). At first, the generating functions approach seemed to be inapplicable for the case of long-range interactions (Chen, 1987). Later studies demonstrated that it can be applied to long-range cooperativity (Kong, 2006; Lando and Nechipurenko, 2008). However, it is difficult to implement if more than one type of large proteins exists in the system (Chen, 1990).

### 2.4. Recurrent relations method

The idea of the recurrent relations method is that a partition function $Z_N$ of a DNA lattice of length $N$ can be expressed in terms of the partition function of a smaller lattice (Gurskii and Zasedatelev, 1978). In particular, the following recurrent expressions hold for non-cooperative protein binding (Nechipurenko *et al.*, 2005):

$$Z_N = Z_{N-1} + K_{N-m+1} c_0 Z_{N-m}; \ Z_N = 1 \ if \ N < m; \ Z_m = K_1 c_0 + 1 \qquad (8)$$

For nonspecific protein binding to a DNA of infinite length, Eq. (8) converges to the McGhee-von Hippel expression (Eq. 5). More general recurrent relations including Eq. 8 as a specific case have been derived for sequence-specific binding with long-range interactions by Gurskii and Zasedatelev (Gurskii and Zasedatelev, 1978). Many dynamic programming approaches in computational biology use an approach similar to that of Gurskii and Zasedatelev: The partition function is derived from a recursive series of simpler calculations. Recent examples include the COMPETE algorithm (Wasson and Hartemink, 2009), which provides a fast way to calculate genome-wide binding of several protein species at the cost of restricting the model only to competitive binding, the GOMER algorithm (Granek and Clarke, 2005), which allows treating long-range interactions between a protein and a DNA promoter but not between two DNA-bound proteins, the algorithm developed by Segal and co-workers for competitive binding of TFs and nucleosomes (Raveh-Sadka *et al.*, 2009; Segal *et al.*, 2006), as well as algorithms to describe protein-protein interactions (Laurila *et al.*, 2009; Lubliner and Segal, 2009).

### 2.5. Transfer matrix method

The transfer matrix formalism was initially used to obtain analytical expressions for simple homopolymer systems (Hill, 1957). However, it allows treating DNA sequence-specificity as well (Crothers, 1968; Gurskii *et al.*, 1972). The method is based on the construction of transfer matrices (weight matrices) for each DNA unit. Each transfer matrix element $Q_n(i, j)$ contains the probabilities to find the lattice unit $n$ in a state $i$ provided the unit $n+1$ is in state $j$. Prohibited combinations of states are characterized by zero weights. The multiplication of all matrices according to the DNA sequence gives the partition function, and its corresponding derivatives yield the probabilities of all binding events (Gurskii *et al.*, 1972). The algorithm for matrix construction depends on the model. In Table 2, the transfer matrix is depicted for the DNA binding of a protein that occupies $m$ DNA units. Each DNA unit may be either bound by a protein or not. Bound states may be subdivided into *m* microstates depending on the unit's position under the protein. Unbound states may be subdivided into three microstates: at the DNA left and right ends, and far from the DNA ends, between bound proteins. The latter states are entropically distinguishable (Teif *et al.*, 2008). Note that transfer matrices determined by Table 2 are very sparse (zero elements are shaded).

**Table 2. A transfer matrix for a model with a single protein type reversible binding DNA**

| State # (unit i \ unit i+1) | 1 | 2 | 3 | … | *m* | *m* + 1 | *m* + 2 | *m* + 3 |
|---|---|---|---|---|---|---|---|---|
| 1 (bound) | | $c_0 K_n$ | | … | | | | |
| 2 (bound) | | | 1 | … | | | | |
| 3 (bound) | | | | … | | | | |
| … | … | … | … | … | … | … | … | … |
| *m* - 1 (bound) | | | | … | 1 | | | |
| *m* (bound) | 1 | | | … | | | 1 | 1 |
| *m* + 1 (free left end) | 1 | | | … | | 1 | 1 | |
| *m* + 2 (free right end) | | | | … | | | 1 | |
| *m* + 3 (free inside) | 1 | | | … | | | | 1 |

Previous studies have shown that the transfer matrix method allows treating sequence-specificity (Crothers, 1968), long-range interactions between bound proteins (Chen, 1987), binding of many protein types (Akhrem *et al.*, 1985; Chen, 2004), double-helical asymmetry of the DNA (Di Cera and Kong, 1996), multilayer protein assembly associated with DNA looping (Teif, 2007) and the competition with nucleosomes (Teif and Rippe, 2009). In addition, this method allows an exact solution of complicated boundary conditions, which is nontrivial at short regulatory DNA regions (Teif, 2010).

**3. Lattice models for nucleosome arrangements along the DNA**
A first lattice model of nucleosome distributions along the DNA was constructed by Kornberg and Stryer (Kornberg and Stryer, 1988). Consider a long array of nucleosomes on the DNA. The number of nucleosomes is fixed, but their positions are variable. Combinatorial laws predict that nucleosomes, which do not have any DNA-sequence preferences, are periodically arranged close to the boundaries of the DNA segment. Such boundary effects are not specific to nucleosomes and apply to any DNA-ligand binding (Epstein, 1978; Di Cera and Phillipson, 1996; Flyvbjerg *et al.*, 2006). Recent experimental studies (Mavrich *et al.*, 2008a; Kharchenko *et al.*, 2008; Cuddapah *et al.*, 2009; Milani *et al.*, 2009) and theoretical considerations (Vaillant *et al.*, 2010; Teif and Rippe, 2009) confirm that this effect is indeed important in nucleosome positioning. For example, the region immediately upstream the transcription start site is usually depleted of nucleosomes. It is not known, whether this depletion is due to nucleosome-excluding sequences or the assembled transcription preinitiation complex. This depleted region forms a barrier, which determines the oscillatory nucleosome positioning pattern decaying at increasing distances (Mavrich *et al.*, 2008a; Kharchenko *et al.*, 2008). Even a stronger barrier is imposed by the insulator binding protein CTCF, which can position about 10 nucleosomes (Cuddapah *et al.*, 2009). Twenty years ago, Kornberg and Stryer proposed that the boundary effects are the first order effect, while "the preferential binding of histones to certain sequences is a second order effect, whose influence upon neighboring nucleosomes is then of third order" (Kornberg and Stryer, 1988). Today the relative contributions of these effects are still under discussion (Jiang and Pugh, 2009; Segal and Widom, 2009a; Radman-Livaja and Rando, 2009; Cairns, 2009), but the importance of statistical positioning is being recognized (Vaillant *et al.*, 2010; Mavrich *et al.*, 2008a; Mavrich *et al.*, 2008b).

A purely entropic model of Kornberg and Stryer was extended by Nechipurenko and coauthors to allow a variable number of nucleosomes on the DNA determined by the thermodynamic equilibrium as in the conventional DNA-ligand binding (Nechipurenko and Vol'kenshtein, 1986; Iovanovich and Nechipurenko, 1990; Nechipurenko, 1988). This has lead to the introduction of energetic parameters – the sequence-dependent nucleosome binding constant, the effective nucleosome concentration and the nucleosome-nucleosome interaction constant. In these models, the parameters could be chosen to obtain an average nucleosome-nucleosome distance equal to the linker length fixed in the model of Kornberg and Stryer. Similar approaches are now actively used to analyze genome-wide nucleosome positioning (Teif and Rippe, 2009; Segal *et al.*, 2006; Schwab *et al.*, 2008; Chevereau *et al.*, 2009; Ioshikhes *et al.*, 2006) (Figure 1).

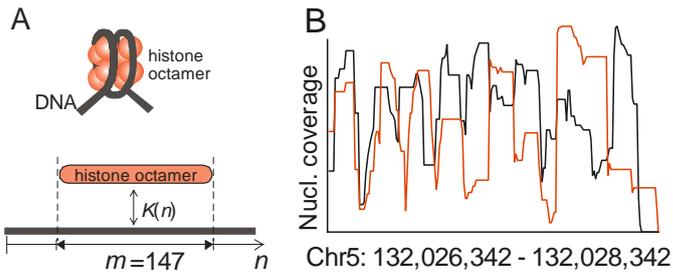

Figure 1. A) The nucleosome represented as a single ligand binding DNA and covering $m = 147$ base pairs upon binding. B) A nucleosome map calculated within this model for a 2000 bp genomic region of resting (black lines) and activated (red lines) human CD4$^+$ T cells (Teif and Rippe, 2009).

In addition to the concepts of boundary- and sequence-determined positioning, a concept of concentration-dependent nucleosome switches has been put forward. Changes in the concentration of histones (or activity of histone-remodeling complexes) can lead to the abrupt switching between different patterns of the nucleosome arrangement (Schwab *et al.*, 2008; Teif and Rippe, 2009). Bistable patterns of nucleosome positioning have indeed been reported for many intragenic regions (Vaillant *et al.*, 2010; Chevereau *et al.*, 2009). Concentration-dependent nucleosome switches are only possible if the total number of nucleosomes on the DNA is not fixed. In general, one should be aware that there are two principally different statistical-mechanical approaches to describe nucleosome arrangement on a one-dimensional DNA lattice. The first approach is based on the Ising model, as exemplified by the McGhee-von Hippel formalism for ligand-DNA binding (McGhee and von Hippel, 1974). In this model, the concentration of ligands in the solution is fixed, while the number of ligands on the DNA is not fixed, and is determined by the thermodynamic equilibrium. The second approach is based on the so-called Tonks gas model as applied to DNA-ligand binding (Woodbury, 1981). This model is based on the assumption that a fixed number of ligands is freely sliding along the DNA. The latter assumption was used by Kornberg and Stryer (Kornberg and Stryer, 1988), while other works have used the assumption of a variable number of nucleosomes (Iovanovich and Nechipurenko, 1990; Schwab *et al.*, 2008; Segal *et al.*, 2006)). In the limit of saturating concentrations (as is the case for the nucleosomes) these two approaches yield significantly different results (Woodbury, 1981; Teif and Rippe, 2009).

*3.1. Interaction energies and concentrations*
Binding energy and concentration are the two parameters central for lattice models of nucleosome chains. One way to determine nucleosome formation energies at physiological ionic strength is via a stepwise dilution of nucleosomes down to the concentrations at which the histone octamer dissociates from the DNA. An apparent dissociation constant can be determined by assuming the equilibrium between the complete nucleosome and the DNA/subnucleosomal species (Rippe *et al.*, 2008), which strongly depends on the ionic properties of the solution. From the early studies a value of $K_d = 3$ nM ($\Delta G = 11.5$ kcal/mol) was derived for histone binding to the bulk mice DNA in a 0.15 M NaCl buffer (Cotton and Hamkalo, 1981). A related approach was used to study nucleosome positioning at a strong positioning sequence from 5S rDNA, which gave $K_d = 0.2$ nM ($\Delta G = 13.2$ kcal/mol) in the presence of 0.15 M NaCl (Gottesfeld and Luger, 2001). However, it is noted that these types of experiments do not yield the true equilibrium dissociation constant since free histones tend to aggregate irreversibly when dissociated from the nucleosome. *In vivo* nucleosome assembly via a complex of histones with histone chaperones such as NAP1 and Asf1 prevents unspecific interactions and aggregation. NAP1 binds linker histone H1, the H2A·H2B dimer and the H3·H4 histone with the same affinity of approximately 10 kcal/mol (Rippe *et al.*, 2008). Histones can be released from NAP1 via competitive binding to the DNA. It is estimated that the energy of binding of the last H2A·H2B dimer to a histone hexasome to form the nucleosome is around -1.5 kcal/mol in the presence of NAP1, while it is around -11.7 kcal/mol in the absence of histone chaperones (Mazurkiewicz *et al.*, 2006; Rippe *et al.*, 2008). The uncertainty of the absolute energies of histone-DNA interactions is an important issue, which is usually neglected in current computational approaches. These operate with *relative* changes of the binding affinities of the nucleosome core for different DNA sequences. This energy difference can vary from zero to -2.4 kcal/mol depending on the DNA sequence *in vivo*, and up to -4.1 kcal/mol for artificial octamer binding sequences *in vitro* (Thåström *et al.*, 2004).

Although the importance of the genomic sequence to nucleosome positioning has been hypothesized quite a long time ago (Trifonov, 1980), systematic calculations of nucleosome positioning on the genomic

DNA started only recently, when a large enough pool of nucleosome-binding affinities was collected (Ioshikhes *et al.*, 2006; Segal *et al.*, 2006). As a result of these studies, not only nucleosome sites at known positions were recovered, but also the probabilities to find nucleosomes at many unknown positions were predicted and verified later. Since the experimental nucleosomal DNA pool used as a computer-training set is increasing very fast, the brute-force bioinformatics approach works pretty well, even without knowing physical details about molecular interactions underlying observed nucleosome distributions. However, not all nucleosome sites can be predicted from the DNA sequence (Stein *et al.*, 2009; Zhang *et al.*, 2009b), and for those which can be predicted, *in vivo* occupancies are not always recovered (Morozov *et al.*, 2009; Teif and Rippe, 2009). This might be explained by the fact that experimental nucleosome patterns are obtained from whole-genome sequencing of DNA isolated by in situ digestion with micrococcal nuclease in the environment of the cell nucleus. Therefore they may reflect to some extent contributions from the higher-order chromatin structures, competing proteins and chromatin remodeling complexes (Stein *et al.*, 2009; Zhang *et al.*, 2009b).

Using high-resolution imaging techniques such as atomic force microscopy (AFM), it is possible to track individual nucleosomes on a given DNA *in vitro* (Bash *et al.*, 2001). A lattice model to analyze experimental nucleosome-positioning data was developed by Solis and coauthors (Solis *et al.*, 2004). The authors determined the number of nucleosomes reconstituted by salt dialysis on DNA segments of known length containing multiple 5S rDNA sequences. In addition to energetic and entropic effects considered in the previous models, this model introduced nucleosome-nucleosome interactions as virial coefficients in the frame of the perturbation theory. From this, the energy of histone octamer and nucleosome-nucleosome interactions were estimated. It appeared that both parameters were dependent on the acetylation state of the nucleosomes. For all studied DNA templates, histone acetylation decreased the affinity of core-particles to the DNA and eliminated nucleosome-nucleosome attractions (Solis *et al.*, 2007). The energies of nucleosome-nucleosome interaction were estimated as -0.25 kcal/mol (attraction) for unacetylated nucleosomes and +0.15 kcal/mol (repulsion) for acetylated nucleosomes. This may partly explain why acetylated nucleosomes are often found at actively transcribed regions – it is easier to remove them. However, it is not clear what is the relevance of this effect in comparison with the action of acetylated histone tails as recruitment tags for remodelers (Choi and Howe, 2009). In addition to tail-mediated nucleosome interactions, there are other levels of interactions mediated by linker histones H1 and stacking interactions of the nucleosomes regularly packed in the chromatin fiber (Kruithof *et al.*, 2009). Single-molecule experiments using optical tweezers suggest a total energy of about 2 kcal/mol for each nucleosome-nucleosome contact (Cui and Bustamante, 2000).

Another important parameter for the calculation of the histone octamer occupancy is the effective concentration (or activity) of the histone octamer in the environment of the nucleus. As already pointed out, the stepwise assembly of histone dimer particles mediated by histone chaperones precludes a direct measurement of this parameter. Let us make a back-of-the-envelope calculation of the histone octamer concentration. The average nucleosome repeat length, *NRL,* is about 200 bp in human cells (other organisms are different (van Holde, 1989)). One nucleosome covers $m$ = 147 bp. Therefore the average degree of DNA coverage is given as $m$ / $NRL \approx 0.75$, which results in a degree of binding $\nu = 0.75 / 147 \approx 0.0051$ (nucleosomes per bp). Substituting these values of $m$ and $\nu$ in the McGhee-von Hippel equation for nonspecific protein-DNA binding (Eq. 5), we get $Kc_0 \approx 0.4$. The absolute value of the concentration $c_0$ depends on the absolute value of the binding constant $K$, which is very sensitive to the experimental conditions, as described above. On the other hand, the value of the product $Kc_0$ gives an idea of the nucleosome behavior in all competitive binding processes. Current bioinformatics software usually treats $c_0$ and $Kc_0$ as unknown scaling parameters, equal to unit by default (Kaplan *et al.*, 2009). According to our estimate above, this default approximation is surprisingly not far from reality. However, when additional features such as the competition with transcription factors, activity of histone chaperones and chromatin remodelers need to be considered (as follows later) one should be more careful with the choice of parameters $K$ and $c_0$ for a more faithful representation of specific experimentally studied systems.

*3.2. Partial assembly of nucleosome particles*
It is well known that the nucleosome is a dynamic structure, which has several intermediate states. It can "breathe" by partially unwrapping the DNA or exchanging histones with the solution. This allows for a transient access of transcription factors to intra- and internucleosomal DNA segments (Shlyakhtenko *et al.*, 2009; Poirier *et al.*, 2008; Poirier *et al.*, 2009; Gansen *et al.*, 2009; Bucceri *et al.*, 2006; Li *et al.*, 2005; Koopmans *et al.*, 2009; Anderson *et al.*, 2002; Anderson and Widom, 2000). The nucleosome particle has a histone octamer core that is organized into four dimers (two H2A·H2B and two H3·H4) Dissociation of these dimers from the nucleosome leads to the formation of subnucleosomal particles (Zlatanova *et al.*, 2009; Gansen *et al.*, 2009). Thus, one has to be aware that the histone octamer is not a single entity and is not a usual ligand.

Nucleosome dissociation can be described by a lattice model as depicted in Figure 2, in which four ligand types are introduced. Although only two types of physically distinguishable subnucleosomal particles H2A·H2B and H3·H4 exist, we need $f = 4$ ligand types to characterize the *f*-mer assembly model (Teif, 2007). For example, a combination of subunits 1-2-3-4 characterizes a standard nucleosome, while combinations 1-2-3 and 2-3-4 characterize a hexasome (a nucleosome lacking one H2A·H2B subunit), and a combination 2-3 characterizes a tetrasome (a nucleosome lacking two H2A·H2B subunits). A new nucleosome can start after an existing one, so the combination 4-1 is allowed. The combinations of subnucleosomal subunits other than the order of 1-2-3-4 are prohibited. A transcription factor TF is defined as a fifth ligand type. The ligand can precede or follow a nucleosome, so the contacts 5-1 and 1-5 are allowed. To make the nucleosome a preferred assembly unit, the contacts 1-2, 2-3 and 3-4 are given high statistical weights, while the other allowed contacts are characterized by the unit weights. Due to the fact that subunits 2 and 3 have two cooperative contacts, they represent the most stable part of the nucleosome core in comparison with the subunits 1 and 4. Obviously such a simple model cannot capture all complex interactions between the nucleosome subunits, but it allows predicting the main feature: during the competitive binding, it is easier to displace either subunit 1 or 4 denoting H2A·H2B. This feature is in accordance with available experimental data (Mazurkiewicz *et al.*, 2006; Zlatanova *et al.*, 2009).

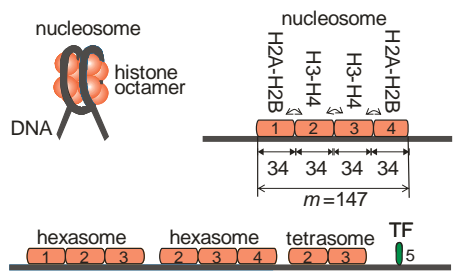

Figure 2. Splitting the nucleosome into subunits. A lattice model treating a single nucleosome core particle as four ligand species ($f = 4$).

*3.3. Nucleosome sliding and ATP dependent chromatin remodelers*
In addition to simple dissociation-reassociation reactions, the nucleosome can slide along the DNA without being disassembled. Nucleosome sliding may happen spontaneously with a low probability (Pennings *et al.*, 1991), or be directed by energy-consuming molecular motors, so-called chromatin remodelers (Whitehouse and Tsukiyama, 2006). Nucleosome sliding is achieved by breaking one or several out of ~14 contacts keeping the DNA and the histone core together in the nucleosome (Längst and Becker, 2004). For example, partial unwrapping of a small segment of the intranucleosomal DNA (e.g. 10-50 bp) can form a loop, which is subsequently propagated around the histone octamer protein core and thus repositions the nucleosome (Schiessel *et al.*, 2001; Längst and Becker, 2004; Chou, 2007; Mollazadeh-Beidokhti *et al.*, 2009). Nucleosomes can also invade territories occupied by other nucleosomes (Engeholm *et al.*, 2009), overlapping with each other and changing the canonical nucleosome structure (Sims *et al.*, 2008).

Chromatin remodelers deviate from the concept of standard "ligands" even further. Unlike usual proteins binding DNA at a thermodynamic equilibrium, remodelers are energy-dependent molecular motors. While their binding/unbinding to the nucleosomes might be still described by simple equilibrium models, the final nucleosome repositioning seems to be more complex, depending on the context, epigenetic modifications and DNA sequence (Rippe *et al.*, 2007). A systematic description of the remodeler action has been proposed based on the concept of remodeler rules (Teif and Rippe, 2009) (Figure 3).

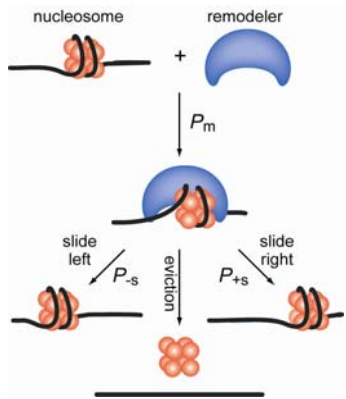

Figure 3. An iterative lattice model to redistribude nucleosomes according to the remodeler rules. Sequence- and context-dependent probabilities describe nucleosome sliding by a step *s* to the left/right or a complete nucleosome eviction.

Initial nucleosome positioning in the absence of remodelers can be calculated using the standard lattice model in Figure 1 assuming that nucleosome positions are determined only by the intrinsic affinities. The effect of chromatin remodelers can be then calculated by iteratively redistributing the nucleosomes according to the remodeler rules until a new steady state is reached. The rule is a probability for a given remodeler to move a given nucleosome from position *n* by a step *s* to a new position *n*+*s* or *n*-*s*. Three major classes of remodeler activities have been identified from this analysis: (i) the establishment of regular nucleosome spacing in the vicinity of a strong positioning signal acting as a boundary, (ii) the enrichment/depletion of nucleosomes through amplification of intrinsic DNA sequence encoded signals, and (iii) the removal of nucleosomes from high-affinity binding sites. The computational study has showed that both the remodeler action and intrinsic nucleosome affinities are important for nucleosome positioning *in vivo* (Teif and Rippe, 2009). A next step should be to derive systematically such rules from the analysis of *in vitro* single-nucleosome experiments and/or genome-wide sequencing. First attempts to solve this task already started (Le *et al.*, 2009), but a successful solution would require more efforts.

*3.4 Linker histones and other architectural proteins*
Transcription factors and architectural proteins can be also considered as ligands, which can reversible bind and unbind DNA. In a system of one non-interacting DNA binder covering *m* DNA units, each DNA unit may be in ~ *m* states (Table 2). In a system of *f* types of interacting DNA binders, each DNA unit may be in ~ $\Sigma(m_g + V_g)$ states, where $m_g$ is the length of a protein of type *g*, $V_g$ is the maximum interaction distance for *g*-type proteins, $g = 1,…,f$ (Teif, 2007). In this model the histone octamer can be considered as one ligand type covering *m* = 147 bp, linker histones as another ligand type covering ~20 bp, and non-histone chromosomal proteins or transcription factors as additional ligand types. H1 and typical transcription factors cover 10-20 bp, while Pol II covers around 30-50 bp. Different thermodynamic features could be assigned to a certain type of nucleosome that for example contains the H2A.Z histone variant instead of the canonical H2A histone.

For a description of linker histone binding, a lattice model was developed, in which each nucleosome provided a single site for nonspecific H1 binding (Ishii, 2000) (Figure 4A). This model predicted that cooperative H1-H1 interactions could lead to the propagation of chromatin "opening" characterized by the loss of H1 histones. The correlation length, defined as a distance over which the presence or absence of H1 proteins affects the binding of new linker histones, was reported to be up to several hundreds of nucleosomes if contact H1-H1 interactions were characterized by the energy 7.5 kcal/mol, and up to several dozens of nucleosomes for the energy of 5 kcal/mol. Analogous models were constructed for a highly abundant non-histone chromatin protein HMGB1, which can compete with H1 and cooperatively assemble on the DNA (Teif *et al.*, 2002). However, later experimental data did not support such strong H1-H1 interactions (Mamoon *et al.*, 2005). On the other hand, H1-DNA binding experiments suggest that H1 binding cannot be fitted by the McGhee-von Hippel type of cooperativity, while it can be described by the two state model with the H1 binding constant for H1-saturated "closed" chromatin domains being 10 times higher than for H1-depleted "open" domains (Mamoon *et al.*, 2005) (Figure 4B). In analogy, a two-state model with different affinities for TF binding to the free and nucleosomal DNA was considered recently (Mirny, 2009). As in the case with H1, this model yields cooperative titration curves that are similar to the two-state models for ligand-induced DNA aggregation (Saroff, 1991), conformational transitions (Poland, 2001) and condensation (Teif, 2005).

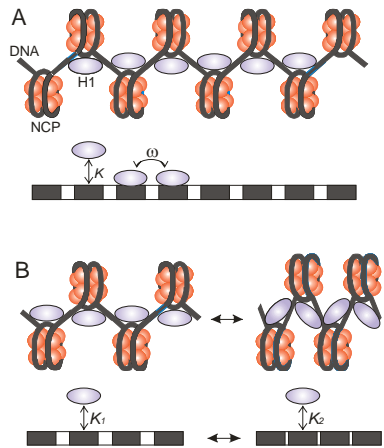

Figure 4. Lattice models to incorporate linker histone proteins. A) Each nucleosome represents one binding site for H1; H1 linker histones interact with each other by contact cooperativity. B) H1 histones do not interact with each other; the cooperativity of binding arises due to different binding constants for H1-saturated and H1-depleted chromatin, which represent different phases.

Additional lattice models are required at higher levels of chromatin compaction. For example, nucleosomal DNA is further compacted with the help of HP1 proteins, each nucleosome representing a binding site recognized by HP1 dimers. While the exact stoichiometry of these interactions is not known, experiments demonstrate the existence of at least three classes of binding sites with different HP1 affinity in chromatin. These differences might reflect the methylation status of histone H3 lysine 9 and/or binding of a HP1 dimer to one or two nucleosomes (Müller *et al.*, 2009). Because of the three-dimensional nature of the chromatin fiber, the two nucleosomes connected by an HP1 dimer are close to each other in 3D, but are not the nearest-neighbors along the 1D DNA lattice (Figure 5). Thus binding/unbinding of HP1 might significantly alter the compaction of the chromatin fiber. Indeed, different concentrations of HP1 have been found in heterochromatin and euchromatin.

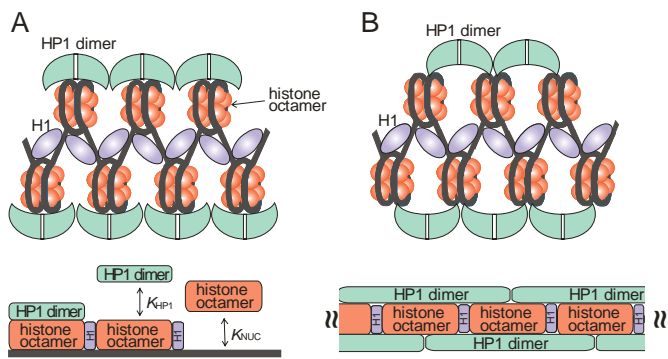

Figure 5. Heterochromatin protein HP1 binds nucleosomal DNA with different affinities, which may affect transitions between heterochromatin and euchromatin. A) One HP1 dimer binds one nucleosome. B) HP1 dimers can connect nucleosomes which are close to each other in 3D and separated by larger distances along 1D DNA lattice. Binding stoichiometry is the same as in (A).

**4. Transcription factor binding to chromatin**

Although a nucleosome can partially unwrapped to allow TF binding (Shlyakhtenko *et al.*, 2009; Poirier *et al.*, 2009), nucleosomes and TF binding to the DNA may be mutually exclusive (Svaren *et al.*, 1994). In the latter case, a nucleosome pre-assembled at a given site cannot be displaced by TF unless chaperones or remodelers mediate the translocation of the histone octamer to a different position or evict the nucleosome. Also, the reverse situation might occur: genomic regions preferentially occluded by transcription factors or Pol II exclude nucleosomes by kinetic rather than equilibrium competition scenarios. Experimentally, it is found that the promoters of transcriptionally active genes are usually depleted from nucleosomes, while the promoters of transcriptionally inactive genes are occluded by the nucleosomes and being remodeled in the process of gene activation (Schones *et al.*, 2008; Kaplan *et al.*, 2009). Many promoters possess both a nucleosome-exclusion signal (such as the rigid poly(dA:dT) repeats) and a TF- or RNAP-positioning signals such as TATA boxes (Zhang *et al.*, 2009a; Radman-Livaja and Rando, 2009; Segal and Widom, 2009b). Thus, it is not clear whether the nucleosome is depleted upstream of TSS at the active promoters due to the inherently low affinity of the histone octamer to the underlying DNA sequence (Segal *et al.*, 2006), or due to the displacement by transcription factors (Morozov *et al.*, 2009) combined with the action of chromatin remodelers (Teif and Rippe, 2009).

*4.1. TF binding to the nucleosomal DNA*
The lattice models shown in Figures 1A and 2 allow treating two types of TF-nucleosome competition. In the simplest case the nucleosome behaves as a single entity (Figure 1A). In the improved model depicted in Figure 2, the nucleosome is allowed to partially disassemble into smaller particles and TFs are allowed to penetrate into the nucleosome. The latter process is energetically unfavorable, but possible. Figure 6 shows results of calculations according to these models. In the absence of nucleosomes, TF is strongly bound to its specific binding site; adding nonspecifically-binding nucleosomes decreases the apparent TF binding constant by several orders of magnitude. Interestingly, when the ligand is allowed to penetrate inside the nucleosome, the binding curve is shifted to increase the apparent TF binding constant. Thus splitting the nucleosome into subnucleosomal particles facilitates TF access to the nucleosomal DNA.

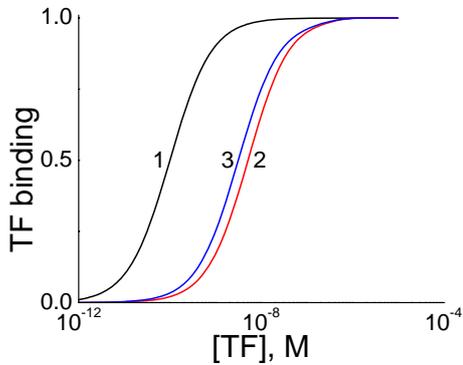

Figure 6. The probability of TF binding to its binding site $n = 1000$ on a DNA lattice on 3100 nucleotides. 1 – TF in the absence of nucleosomes. 2 – TF/NCP competition, NCP is a single entity. 3 – TF can penetrate inside the nucleosome consisting of four subunits (2 × H3-H4 and 2 × H2A-H2B). Contacts between nucleosome subunits are characterized by a cooperativity constant $w = 100$. $c_0$(nuc) = $10^{-7}$ M. $K$(nuc) = $10^9$ M$^{-1}$ for any $n$; $K$(TF, $n = 100$) = $10^{10}$ M$^{-1}$, $K$(TF, $n \neq 100$) = $10^6$ M$^{-1}$

*4.2. TF-TF cooperativity induced by nucleosomes*

Another insightful concept for TF-nucleosome competition is the so called "collaborative competition" (Adams and Workman, 1995; Polach and Widom, 1996). Consider the nonspecific binding of the histone octamer to the DNA in the absence of nucleosome-nucleosome interactions competing with a sequence-specific binding of transcription factors. Let there are two TF binding sites near each other that are both covered by a nucleosome (Figure 7A). Once the DNA segment is freed from the nucleosome by one TF, it is easier for the second TF to bind. Because multiple regulatory DNA sites often occur within a nucleosome-length distance, this cooperativity is to be expected *in vivo*, and has indeed been observed experimentally (Adams and Workman, 1995; Miller and Widom, 2003). It might be explained also by the recruitment of specific remodelers rather than by a simple mechanical competition (Hebbar and Archer, 2007). Figure 7A shows an illustrative map of binding for a DNA region of 3100 bp containing two specific TF binding sites separated by 50 bp. Here, it is assumed that nucleosomes nonspecifically bind the DNA at a high concentration: $K$(NCP)·[NCP] = 100. The nucleosome binding periodicity predicted for this system by Figure 7B is due to the boundary conditions (nucleosome hanging out from this region is prohibited). The periodic nucleosome pattern is disturbed in the vicinity of TF binding sites. This plot shows that for a given set of parameters TFs always partially displace nucleosomes, creating a nucleosome-depleted region.

In order to quantify the nucleosome-induced collaborative effect in terms of the standard McGhee-von Hippel type cooperativity, the logarithmic binding curves ν(ln[TF]) and the Scatchard curves (ν/[TF] as a function of ν) have been calculated in Figures 7C and 7D. Three situations have been considered. In the first case TF binds the DNA in the absence of nucleosomes, which results in a non-cooperative curve. The absence of cooperativity is manifested by the linear Scatchard plot in Figure 7D. In the second case, nucleosomes are added to the system, which decreases the effective TF binding constant, makes the binding curve in Figure 7C slightly steeper, but has almost no detectable curvature in the Scatchard curve in Figure 7D. For comparison, when TF binding is characterized by the standard McGhee-von Hippel cooperativity with $w = 10$ (lower-middle range of typical TF-TF interactions), the resulting Scatchard curve shows a significant curvature characteristic for the binding cooperativity. Thus for a chosen set of parameters, nucleosome-assistant collaborative competition leads only to a moderate cooperativity in comparison with usual protein-protein interactions.

TF-nucleosome competition covers a wide range of effects beyond the two examples above. They can all be described by the corresponding lattice models. Consistent with this, recent papers from several groups consider nucleosomes as competitors to TF binding (Morozov *et al.*, 2009; Teif and Rippe, 2009; Raveh-Sadka *et al.*, 2009; Wasson and Hartemink, 2009), while another possibility is to consider nucleosome-occupancies as static corrections to TF affinities (Gordân *et al.*, 2009). However, the view of nucleosomes as dynamic regulators of gene expression in chromatin allows larger flexibility.

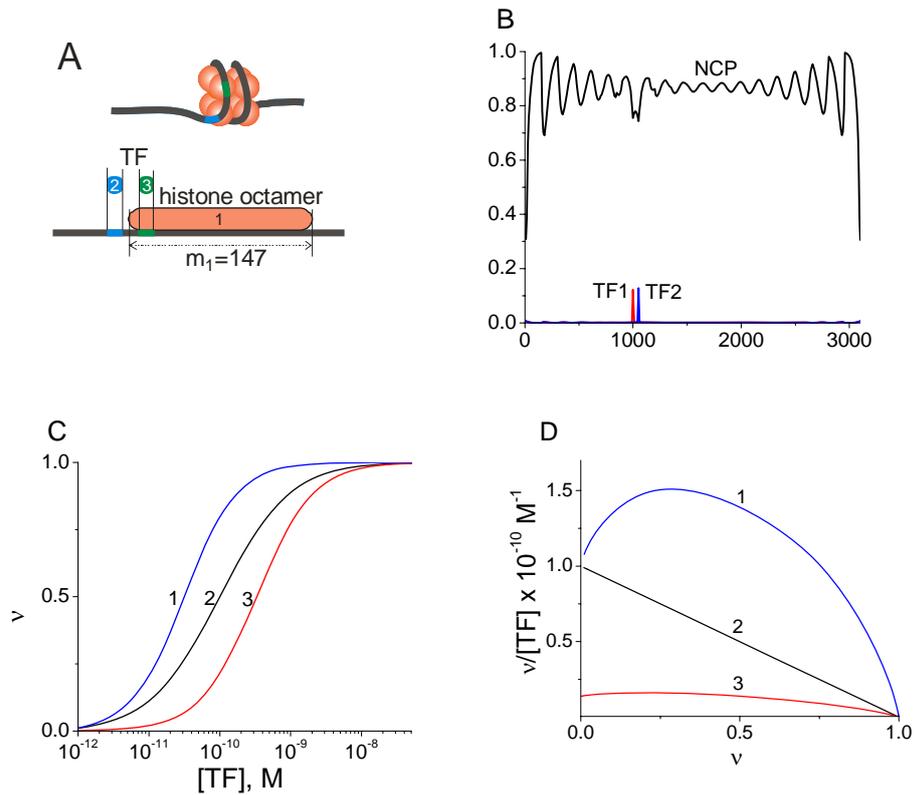

Figure 7. Collaborative competition between transcription factors and nucleosome. A) A scheme of the system: the histone octamer is considered as a single entity, $m$(nuc)=147 bp; transcription factors are characterized by the length $m$(TF)=10 bp; there are two sequence-specific TF binding sites near each other. $c_0$(nuc) = $10^{-7}$ M, $K$(nuc) = $10^9$ M$^{-1}$, $K$(TF, n ≠ 1000, 1050) = $10^7$ M$^{-1}$, $K$(TF, n = 1000, 1050) = $10^{10}$ M$^{-1}$, $c_0$(TF) = $10^{-9}$ M$^{-1}$. C) and D) TF binding sites are separated by 10 bp. TF binding in the presence of nucleosomes (1) is compared to the noncooperative binding in the absence of nucleosomes (2) and the binding characterized by the MvH cooperativity ($w = 10$) in the absence of nucleosomes (3).

## 5. Conclusions and perspectives

We have described a general theoretical framework for applying DNA lattice models to gene regulation in chromatin, including several new models. Different aspects of the nucleosome particle, architectural proteins, chromatin remodeler activity and transcription factor binding can all be described using appropriate lattice models. It is hoped that the approach described here will facilitate the construction of lattice models for other specific problems of protein-DNA interaction in the context of chromatin. One exciting aspect that needs to be addressed by further developments is the presence of covalent epigenetic modifications. These post-translational modifications of the DNA (by methylation) and of the histones (by acetylation, methylation, phosphorylation etc) modulate the binding affinity of chromatin ligands. By serving as anchors for architectural proteins and transcription factors they define gene expression programs of different cell tissues in a manner that is preserved through cell division (Probst *et al.*, 2009). To some extent, epigenetic modifications have already been taken into account in the models discussed above as a correction to the sequence- and context-dependent binding constants $K(n, g)$. By assigning additional states to the elementary unit of the lattice model the modified/unmodified DNA/histone can be described in addition to the bound/unbound states. However, including the propagation of epigenetic pattern would require more extensions to existing lattice models and new approaches. Several approaches of this type have been considered recently albeit in the absence of explicit considerations of protein concentrations and binding affinities (Sneppen *et al.*, 2008; Dodd *et al.*, 2007; David-Rus *et al.*, 2009). Thus, future theoretical and experimental advancement will hopefully fill this gap to yield DNA lattice models that are appropriate for systematic biophysical description of gene regulation in chromatin.

**Acknowledgements**
This work was supported by a fellowship of the German CellNetworks - Cluster of Excellence (EXC81) to VT and by DFG grant Ri 1283/8-1 to KR.